\documentclass[a4paper]{jpconf}
\usepackage{subfig}
\usepackage{wrapfig}
\usepackage{url}
\usepackage{amsmath}
\usepackage{wrapfig}
\usepackage{graphicx}
\usepackage{lineno}
\newcommand{\degC}[1]{\mbox{$#1\,^\circ{\mathrm C}$}}
\begin{document}
\title{Improved Detection of Supernovae with the IceCube Observatory}
\author{Lutz K\"opke for the IceCube Collaboration~\footnote{\protect\url{http://icecube.wisc.edu}}}
\address{}
\begin{abstract}The IceCube neutrino telescope monitors one cubic kilometer of deep Antarctic ice by detecting Cherenkov photons emitted from charged secondaries produced when neutrinos interact in the ice. The geometry of the detector, which comprises a lattice of 5160 photomultipliers, is optimized for the detection of neutrinos above 100 GeV. However, at subfreezing ice temperatures, dark noise rates are low enough that a high flux of MeV neutrinos streaming through the detector may be recognized by a collective rate enhancement in all photomultipliers. This method can be used to search for the signal of core collapse supernovae, providing sensitivity competitive to Mton neutrino detectors to a supernova in our Galaxy. An online data acquisition system dedicated to supernova detection has been running for several years, but its shortcomings include limited sampling frequency and the fact that the burst energy and direction cannot be reconstructed. A recently developed offline data acquisition system allows IceCube to buffer {\em all} registered photons in the detector in case of an alert with low probability to be erroneous. By analyzing such data offline,  a precision determination of the burst onset time and the characteristics of rapidly varying fluxes, as well as estimates of the average neutrino energies may be obtained. For supernovae ending in a black hole, the IceCube data can also be used to determine the direction of the burst. 
\end{abstract}

\section{Introduction}
The IceCube neutrino telescope~\cite{bib:Mainboard,bib:PMT,bib:DetectorPaper,bib:Realtime}, embedded in the clear east antarctic ice sheet, is uniquely suited to monitor our Galaxy for supernovae due to its 1 km$^3$ size. Deployed at depths between 1450 -- 2450 m, the detector is partly shielded from cosmic ray muons; in the inert and \degC{-43} to \degC{-20} cold ice, the dark rates of IceCube's 10-inch diameter photomultipliers average around 540 Hz. While designed to reconstruct interactions of neutrinos with hundreds of GeV energy, such a low dark rate in conjunction with a vast amount of interactions in the instrumented volume allows for the detection of  $ \mathcal{O}(10\,{\rm MeV})$ neutrinos from galactic core collapse supernovae.
The inverse beta process $\bar \nu_\mathrm{e} + \mathrm{p} \rightarrow \mathrm{e^+} + \mathrm{n} $ dominates supernova neutrino interactions in  the ice, leading to positron tracks of about $0.6\, {\rm cm} \cdot E_\nu/ \,{\rm MeV}$ length which radiate $\approx 180\cdot E_\mathrm{e^+}/{\rm MeV}$ Cherenkov photons in the 300 -- 600 nm wavelength range. Due to the approximate $E_\nu^2$ dependence of the cross section at supernova neutrino energies and the linear energy dependence of the positron track length, the light yield per neutrino interaction scales with $E_\nu^3$. Photons travel long distances in the clear ice such that each light sensor effectively monitors several hundred cubic meters. In more than 98\% of the cases, only a single photon from each interaction reaches one of the light sensors because they are vertically (horizontally) separated by roughly 17 m (125 m). The DeepCore subdetector~\cite{bib:DeepCore}, equipped with a denser array of high efficiency photomultipliers, provides higher detection and coincidence probabilities. 

The effect of a high neutrino flux brightening the detector will be clearly seen in the collective rate increase averaged over many sensors. IceCube is sensitive around the clock with 99.7\% uptime and is the most precise detector for analyzing the neutrino light curve of close supernovae~\cite{bib:snpaper}. Indeed, for a supernova at the center of our galaxy, IceCube will remain competitive even in case the proposed Hyper-Kamiokande~\cite{bib:HKletter,bib:HKdesign} detector is realized.

However, the statistical interpretation of the IceCube data does not allow us to reconstruct individual neutrino interactions and estimate their energy, origin and type. Furthermore, the standard supernova data acquisition is based on count rates of individual optical modules stored in 1.6384 ms time bins, which limit the timing accuracy.
Additionally, an artificial deadtime of 250 $\mu$s, incorporated in the data acquisition system  to cut the dark rate in half, has a detrimental effect in the detection capabilities of very close supernovae.

Therefore an improved readout system has been developed to additionally buffer and extract the full photomultiplier raw data stream around supernova candidate triggers (HitSpooling)~\cite{bib:Heereman,bib:HeeremanThesis}. In this paper,  we will introduce the HitSpool system and  discuss the following examples that build on this improvement: 
\begin{itemize}
\item the application of sophisticated techniques to identify noise hits or hits associated with cosmic ray muons that do not trigger the IceCube simple majority trigger~\cite{bib:ICRC2015,bib:HeeremanThesis},
\item the estimation of the average neutrino energy by the use of coincidence probabilities~\cite{bib:Bruijn,bib:Mathieu}, 
\item a directional measurement by internal triangulation in the case a signal changes abruptly and the possibility to measure an absolute neutrino mass measurement in the case of core collapse supernovae ending up as a black hole~\cite{bib:ICRC2015}, and
\item the study of standing wave shock signatures in IceCube and gravitational wave detectors. 
\end{itemize}
Since 2009, IceCube has been sending real-time datagrams to the Supernova Early Warning System (SNEWS)~\cite{bib:antonioli-2004-6} when detecting supernova candidate events. A similar alert system, with substantially lower thresholds to trigger gravitational wave observations, is under preparation~\cite{bib:Leonor,bib:Grimov,bib:Casentini}.

\section{Technical implementation of the HitSpooling system}
\begin{wrapfigure}[17]{o}{0.5\textwidth}
\includegraphics[angle=0,width=0.5\textwidth]{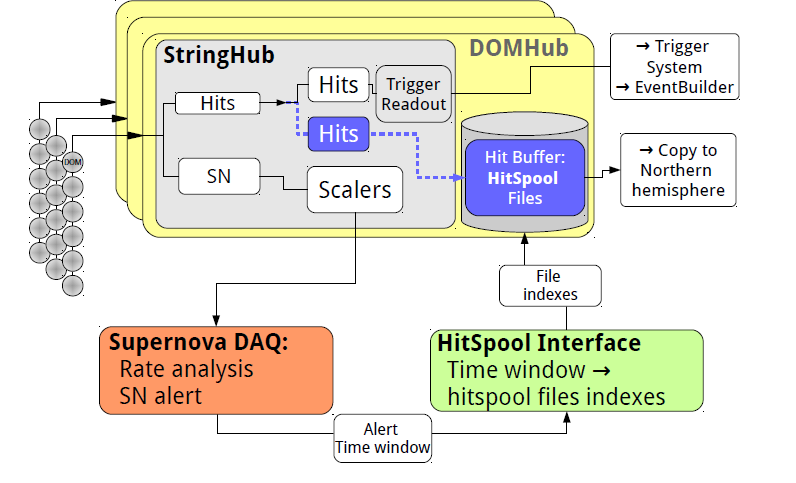}
\caption{\small Block diagram of surface data acquisition components involved in the HitSpool data
stream (blue dashed line). Hits are copied to the DOMHubs hard disk from
where files may be selected and transmitted to the Northern hemisphere via satellite.}
\label{fig:HitSpoolBlock}
\end{wrapfigure}

Fig.~\ref{fig:HitSpoolBlock} contains a logical diagram of the path of data in IceCube and its connection to the supernova data acquisition system. The 5160 digital optical modules (DOMs) in IceCube,
containing photomultipliers and readout electronics, are deployed on 86 so-called strings with 60 DOMs per string. 
The hit stream from each DOM is buffered in a dedicated industrial computer (DOMHub), where the hits are chronologically ordered and packaged by the StringHub component into a form suitable for the application of software-based triggers and event builders. A copy of the raw data is also written to a circular buffer on disk (HitSpool). Old data are overwritten after a spooling cycle of about 72 hours or longer, depending on the storage capacity of the DOMHub.

In the event of a significant transient, a request is sent to the HitSpool system to store hits  within a configurable time interval. For supernova candidates, a period of 30 s before to 60 s after a SNEWS trigger is stored. The supernova system performs HitSpool requests on average twice per month. Such data -- buffered at an early stage of the data acquisition -- will be available in the unlikely case that the data acquisition may fail (i.e. in the case of an extremely close supernova which could exhaust the system). The fully automatic hitspooling system (see data flow scheme in Fig.~\ref{fig:HitSpoolBlock}) has been working reliably for several years. HitSpool processing includes the automatic transfer of data to the North and the subsequent monitoring and pre-processing. 

\section{Examples for recent improvements in IceCube's supernova detection}
In this section, we provide four examples that demonstrate how the HitSpooling capability extends IceCube's physics capabilities either by allowing for the exploitation of coincidences between optical modules or by a much improved temporal precision. 

\subsection{Cosmic ray muon background}\label{sect:cosmic}
\begin{figure*}[ht]
\centering
\subfloat{{\includegraphics[angle=0,width=0.45\textwidth]{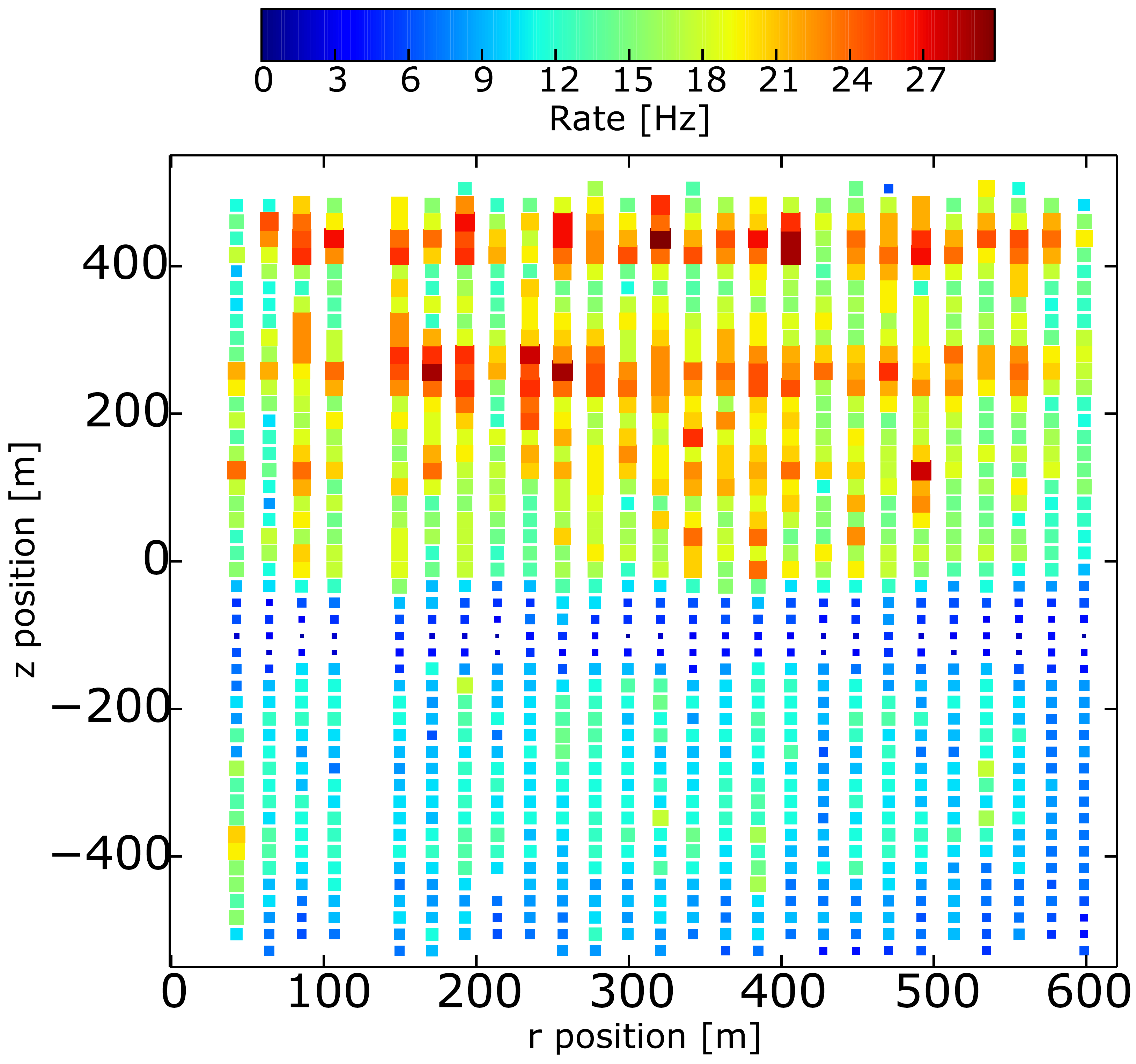}}}
\qquad
\subfloat {{\includegraphics[angle=0,width=0.45\textwidth]{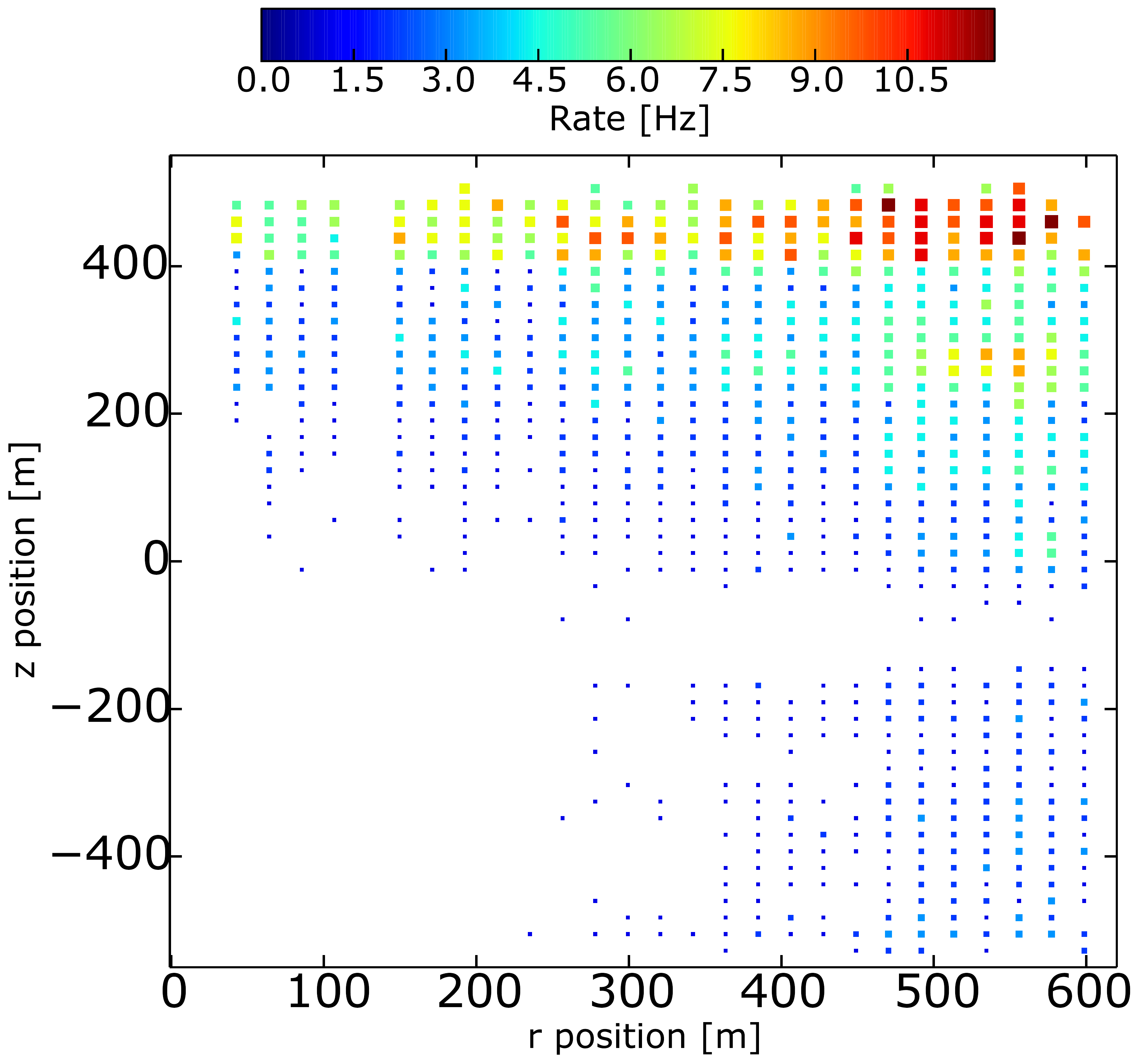}}}
\caption{\small Simulated muon induced hit rates as function of horizontal position and radius w.r.t. to the detector center at 1950 m depth. {\em Left:} all hits associated with muons triggering the detector. {\em Right:} hits associated with muons not meeting any trigger condition.}
\label{fig:PulseDistributions}

\end{figure*}

Atmospheric muons require an energy of $\approx 400$ GeV to reach the detector and $\approx$ 550 GeV to trigger the IceCube 8-fold majority trigger. 
Hits from muon tracks that fail this trigger requirement or that clip the corners of the detector are mostly found
in the upper or outer layers of the detector (see Fig.~\ref{fig:PulseDistributions}).

\begin{figure}[ht]
\includegraphics[angle=0,width=0.46\textwidth]{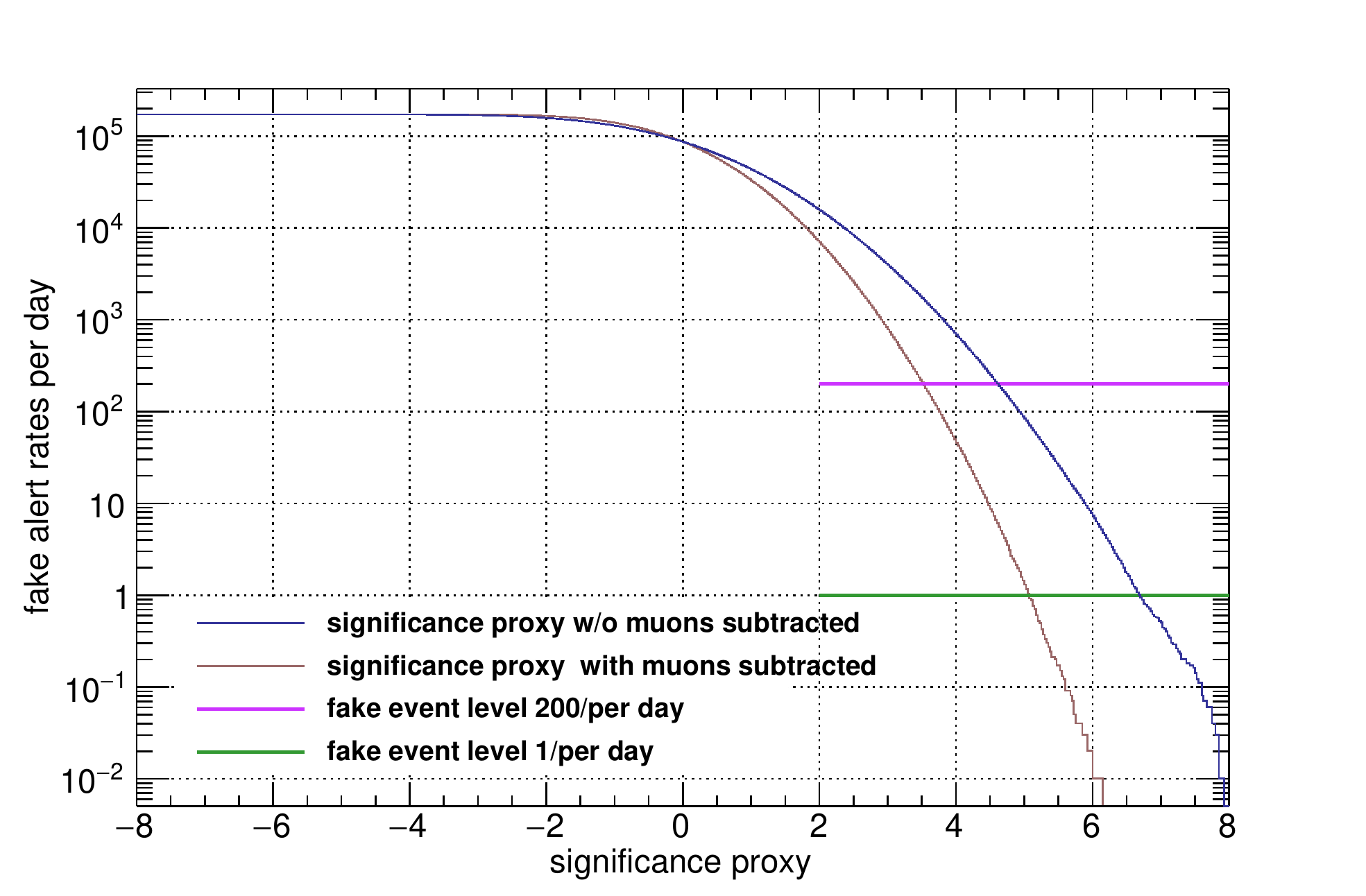}
\includegraphics[angle=0,width=0.47\textwidth]{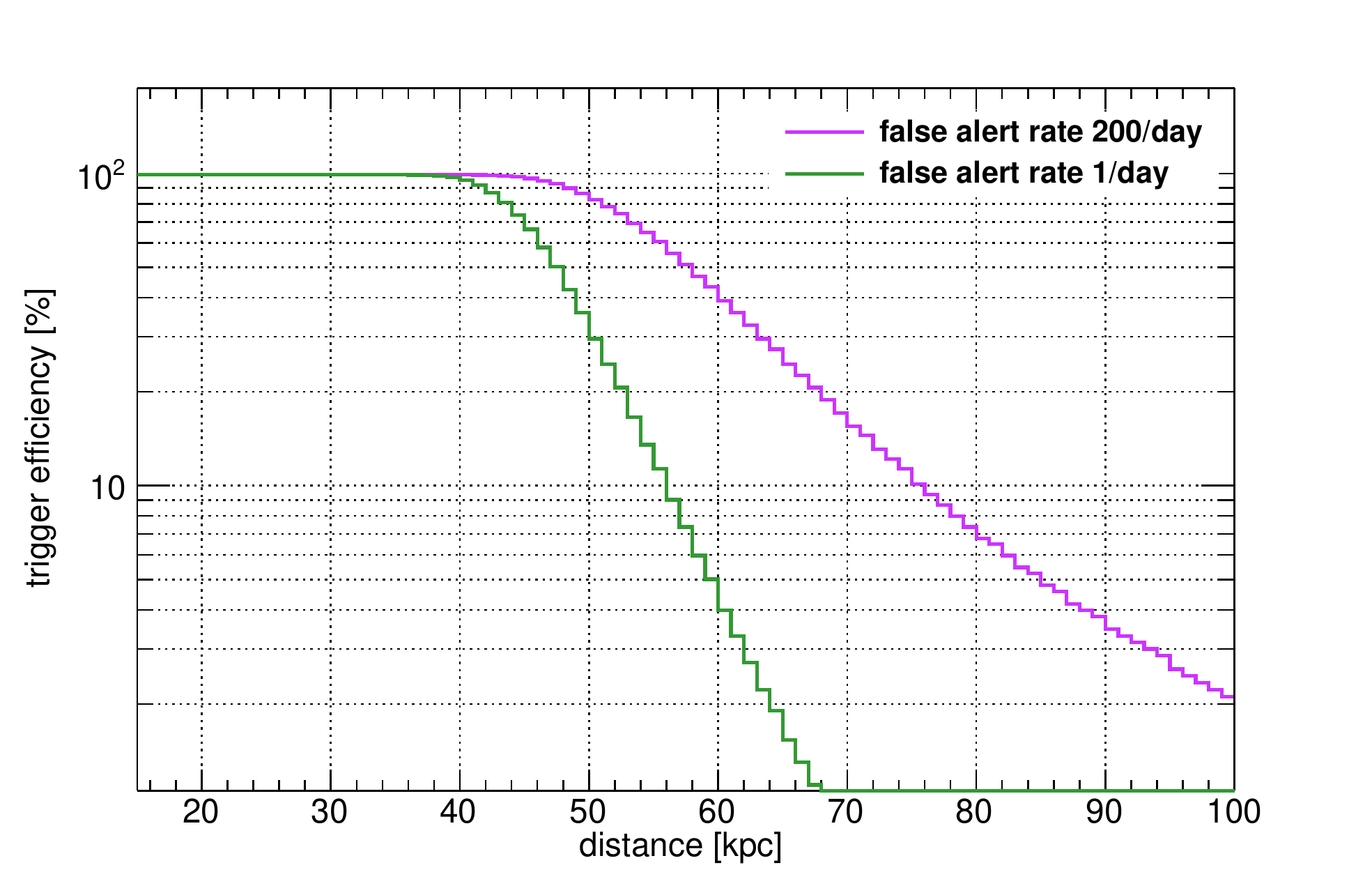}
\caption{Left: Fake alert rate vs. significance proxy; cosmic ray muon subtraction reduces the rate significantly at large significances. Right: trigger efficiency as function of distance for a high alert rate (200 per day in purple) and one alert per day (green).}\label{fig:TriggreEffI3}
\end{figure}

Dust layers in the ice, particularly the one $\approx$ 100 m below the detector center, absorb light. This effect and the limited muon range lead to a depth-dependent cosmic muon induced rate between $\approx$ 3 Hz and $\approx$ 30 Hz. 
The fraction of such hits in each DOM is small compared to the dark rate. However, space and time correlated hits from muons lead to non-Gaussian tails that broaden substantially the significance distribution of the excess count per sensor. 
In fact, 99.9\% of false positive triggers are due to statistical agglomerations of cosmic ray muon induced hits (see Fig.~\ref{fig:TriggreEffI3}). 

To correct for the effect of muons, the number of locally coincident particle induced hits counted by IceCube's majority trigger is transmitted to the supernova data acquisition system and is used to statistically subtract hits associated to cosmic muon induced tracks~\cite{bib:ICRC2015}. The method has allowed us to lower the alarm thresholds and reduce false-positive triggers. The muon rate correction has enabled us to improve the SNEWS alarm efficiency for potential supernovae in the Large Magellanic Cloud from merely 12\% to 82\%, while meeting the SNEWS requirement that alarms are sent with a frequency of less than one alarm per 14 days.

Simulations show that $\approx$50\% of all atmospheric muons crossing the detector produce a simple majority trigger. The remaining sub-threshold muons can be identified using algorithms developed to identify hits clustering in time and space combined with an investigation of the angular multiplicity of tracks formed by each hit pair in the cluster. Thereby it is possible to recover up to 45~\% of all sub-threshold muon hits, limited only by the requirement of at least 4 hits to define a cluster in the algorithms. Applying these techniques to experimental data results in a $\sim$ 3~\% decrease of the total hit rate and reduces the fraction of false positive SNEWS alarms by an additional 45\%~\cite{bib:ICRC2015,bib:HeeremanThesis}. 

\subsection{Determination of the average neutrino energy}
It has been shown~\cite{bib:Bruijn} that the average neutrino energies can be reconstructed from a comparison of single rates with coincident rates in the detector. The probability of photons from a single interaction to reach different optical modules increases with the energy of the neutrino. For each optical module added in coincidence, another factor proportional to the neutrino energy is added. This sensitivity of coincident hits to the neutrino energy has been recognized in~\cite{bib:Mathieu} and was elaborated with a full Monte-Carlo simulation based on Geant4 with custom photon tracking. The uncorrelated background is greatly reduced by the use of a small coincidence gate of 150 ns for IceCube and 100 ns for DeepCore. The correlated background from atmospheric muons has been reduced by a multiplicity cut with further improvements possible (see section~\ref{sect:cosmic}).   

\begin{figure}[t]
\includegraphics[trim=0cm 0cm 0cm 0cm,clip,width=0.44\textwidth]{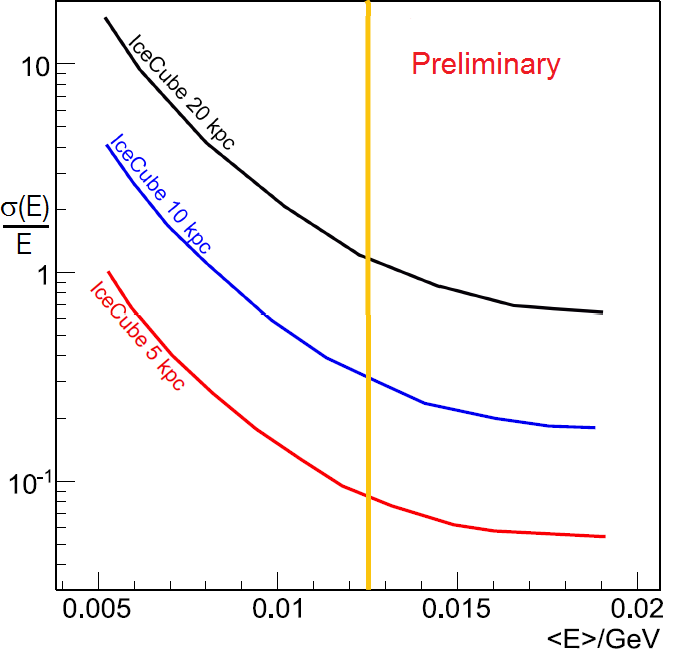}%
\includegraphics[trim=0cm 0cm 0cm 0cm,clip,angle=0,width=0.46\textwidth]{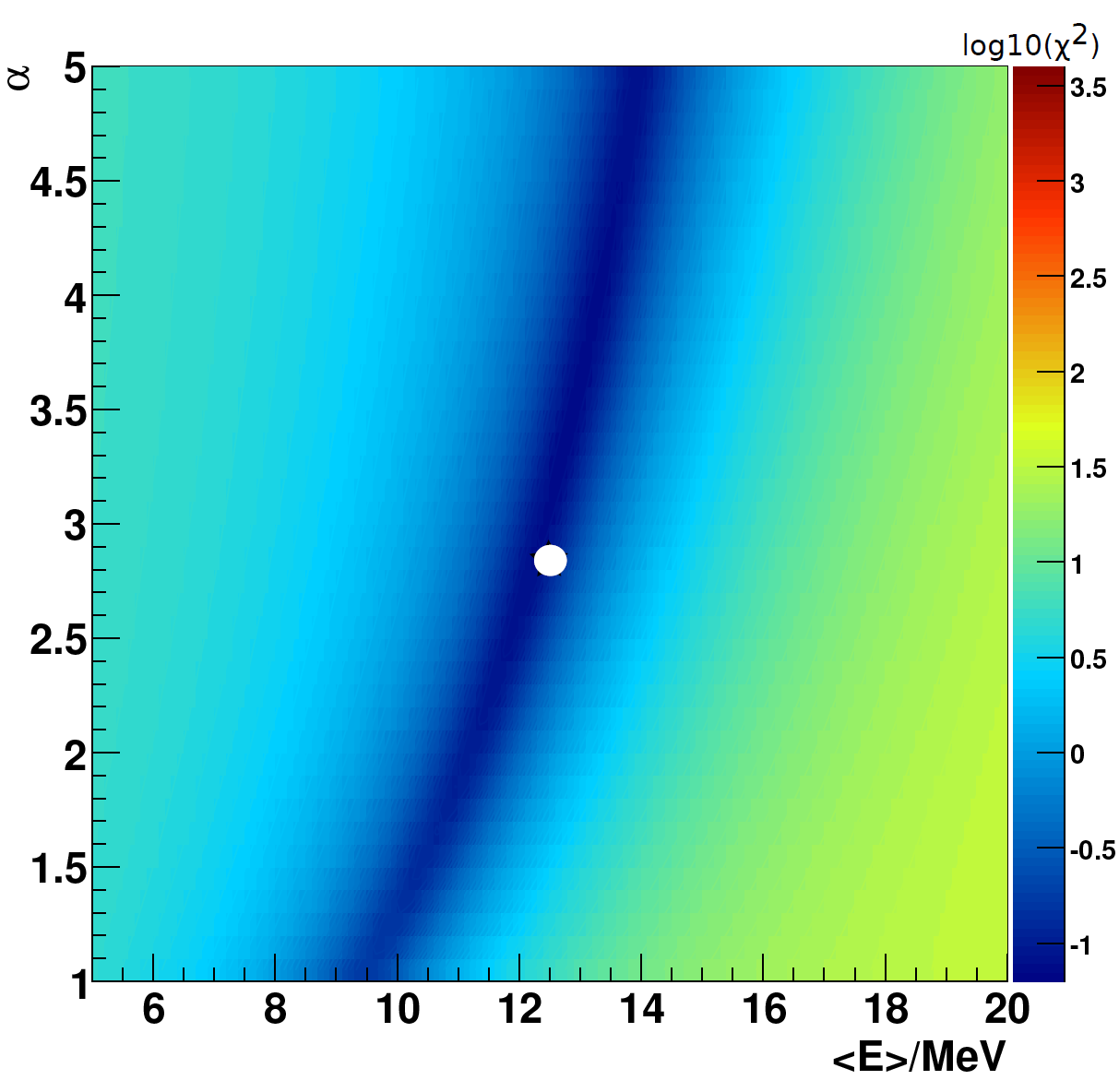}
\caption{\small Left: Comparison of the precision on the determination of the average neutrino energy of a supernova at 20 (black), 10 (blue) and 5 (red) kpc distance. Right: Combined determination of the average neutrino energy and the spectral shape parameter $\alpha$ for IceCube using a $\chi^2$ method. The input value ($\langle E_{\bar\nu_{\rm{e}}}\rangle=12.6$ MeV,
  $\alpha=2.84$) is denoted by a white circle.}
\label{fig:energyres}
\end{figure}

The relative precision of the energy estimate is depicted in Fig.~\ref{fig:energyres} as function of distance and average energy. To be most conservative,
flux, energy and spectral shape were taken from the collapse of an O-Ne-Mg 8.8~$M_\odot$ progenitor star~\cite{bib:Huedepohl}, the lowest mass progenitor
known to undergo a core collapse. Systematic uncertainties have not yet been assessed; however, they are reduced as the ratio of the coincidence rate to the singles rate is used in the method.

Multiple coincident hit conditions show distinct differences in the neutrino energy dependence. These can be used to characterize the
spectral shape of the neutrino emission, in addition to the average energy.  Assuming a neutrino spectral shape parametrized with three
parameters, the luminosity $L_\nu$, the average neutrino energy $\langle E_{\bar\nu_{\rm{e}}}\rangle$ and a shape parameter
$\alpha$~\cite{Keil} and using single, double and triple coincident hit modes in a $\chi^2$ fit, $ \langle E_{\bar\nu_{\rm{e}}}\rangle$
and $\alpha$ can be extracted simultaneously.

A scan of the parameter space for a simulated supernova at 10~kpc distance is shown in Fig.~\ref{fig:energyres} as function of average
energy and $\alpha$ for IceCube. The color scale indicates the log$_{10}$($\chi^2$) value for the parameter values tested. Note
that $\alpha$ and $\langle E_{\bar\nu_{\rm{e}}}\rangle$ are almost degenerate.  

\subsection{Triangulation of the supernova direction and absolute neutrino mass}
For neutrino energies between 10-20 MeV, the direction of the emerging positron is essentially uncorrelated with the incoming neutrino direction. Therefore it is a challenge to measure the supernova direction with IceCube. A triangulation between distant detectors has been proposed~\cite{bib:Triangulation} as one possibility. Another option is to determine the direction from the temporal hit pattern seen in the cubic-kilometer detector when a neutrino wavefront changes its intensity abruptly. The detector crossing lasts only several microseconds, which sets the time-scale for abrupt flux changes to be suitable.

\begin{figure}[h]
\includegraphics[trim=0cm 0cm 0cm 0cm,clip,width=0.45\textwidth]{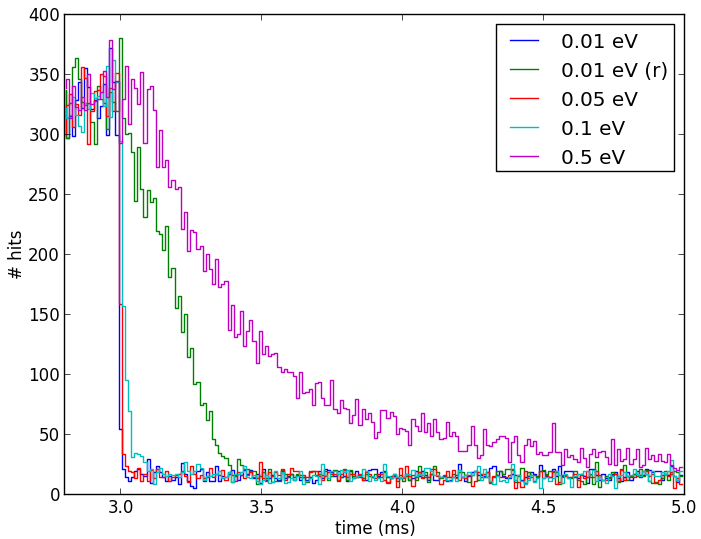}%
\includegraphics[trim=0cm 0cm 0cm 0cm,clip,angle=0,width=0.46\textwidth]{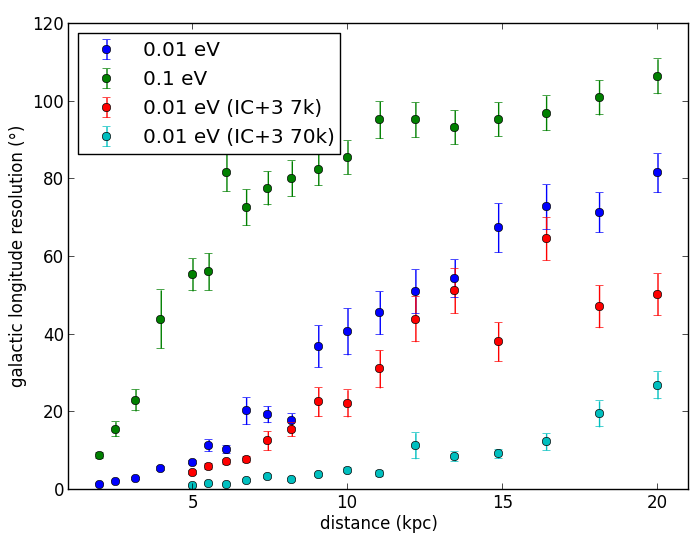}
\caption{\small Left: Effect of neutrino mass on the observed neutrino flux for abrupt black hole creation. The green line shows the effect of a gradual decrease of the neutrino luminosity due to rapidly rotating progenitor (m$_\nu$ = 0.01 eV). Right: Resolution in galactic longitude achievable by triangulation within IceCube's geometry for a supernova with abrupt black hole formation for two assumptions on the mass of the lightest neutrino (0.01 eV and 0.1 eV, respectively) as well as for a hypothetical detector extension with 3 additional IceCube strings deployed symmetrically around IceCube's center at 7 and 70 km distance, respectively.}
\label{fig:angularresolution}
\end{figure}

One example is the formation of a black hole following a core collapse of a supermassive star. In case the protoneutron star forms a black hole, the neutrino flux should cease almost immediately, once the Schwarzschild condition is met. However, since the neutrino energies follow a spectrum, non-vanishing neutrino masses lead to a smearing of the time arrival distribution (see Fig.~\ref{fig:angularresolution}) that can be used to determine the absolute neutrino mass~\cite{bib:Beacom}.
Note that -- in the case of rapidly rotating progenitors -- the neutrino fluxes may also gradually decrease as more and more matter in the star approaches the event horizon and the gravitational redshift becomes extremely strong~\cite{bib:blackhole2} (see Fig.~\ref{fig:angularresolution}).  We first adopt the optimistic scenario that the smearing due to the black hole dynamics is negligible.

With an unbinned likelihood analysis, simulations show that a reasonable directional resolution can be achieved for nearby supernovae and low neutrino masses.  Fig.~\ref{fig:angularresolution} shows the resolution as function of distance assuming rates and energies according to~\cite{bib:blackhole} and an abrupt decrease of the neutrino luminosity at the black hole surface. When using the assumption of a gradual decrease of neutrino luminosity~\cite{bib:blackhole2} instead, one finds that directions can only be determined for very close supernovae.
As the time delay of a neutrino with mass $m_i$ and energy $E_\nu$ at distance $D$, $\Delta t = 0.515 \cdot \frac{{m_i}^2}{\mathrm{eV}^2}\frac{\mathrm{MeV^2}}{E_\nu^2}\cdot \frac{D}{\mathrm 10 \mathrm{kpc}}$,  is proportional to the squared neutrino mass, the method has only little sensitivity for current cosmological neutrino mass limits. However, sterile neutrinos may make an imprint through their mixing with ordinary neutrinos. Neglecting standard neutrino masses and assuming only one sterile neutrino with mass $m_4$, the $\bar{\nu_e}$ flux $F_{\bar{\nu_e}}(E_\nu, t)$ at time $t$ after creation is given by
\begin{eqnarray}
F_{\bar{\nu_e}}(E_\nu, t) &\approx &|U_{e1}|^2 F^0_{\bar{\nu_e}}(E_\nu, t)+ |U_{ei}|^2 F^0_{\bar{\nu_x}}(E_\nu, t)+\widetilde{|U_{e4}|^2 } F^0_{\bar{\nu_x}}\left(E_\nu, t-\frac{D}{2c}(\frac{m_4}{E_\nu}^2)\right)\quad , \nonumber
\end{eqnarray}
with i = 2, 3 for inverted and normal hierarchy, respectively~\cite{bib:Esmaili}. $U_{ei}$ denote the entries in the upper row of the PNMS matrix. Note that $\widetilde{|U_{e4}|^2 }$ should be regarded of as a free parameter since matter and neutrino coherence effects in supernovae, which are hard to predict, may alter the coupling~\cite{bib:Raffelt}. First simulations with an unbinned likelihood fit show that a sterile neutrino mass $m_4=2$eV  can be well reproduced for  
$\widetilde{|U_{e4}|^2 }$ larger than 0.5, for an instantaneous black hole forming core collapse at 10 kpc distance. For this analysis, $ \mathcal{O}(10~\mu$s) timing resolution is essential. 
\subsection{Supernovae and gravitational waves}
While the current sensitivity of Advanced LIGO is limited to supernovae within a few kpc (see e.g.~\cite{bib:Andresen}), technological improvements will enlarge the supernova detection range of gravitational wave detectors to that of current neutrino detectors. Lacking observations, there is a strong variation in modeling the core collapse and the occurrence of asymmetries that lead to gravitational wave emission. In case the progenitor star is rotating~\cite{bib:Ott,bib:Ott2,bib:Yokozawa}, the bounce of the iron core leads to a pronounced spike at bounce time and a subsequent damped ring down with kHz frequencies. 
Generally, non-axisymmetric rotational instabilities such as standing accretion shock instabilities (SASI), will also result in the emergence of a quadrupole mass-moment during the collapse and thereby leave their imprints on the gravitational wave signal.
 
The possibility to search for time correlations between neutrino and gravitational wave data was investigated using recent simulations from~\cite{bib:Andresen,bib:Tamborra}. Fig.~\ref{fig:angularresolution} (left) shows the result of a Monte Carlo simulation, both for the neutrino signal in IceCube (normal hierarchy assumed, distance 3 kpc) and a gravitational wave signal at the source (no noise and detector simulation added). The neutrino signal shows a clear oscillatory pattern  150-250 ms after the burst onset.  At first sight, no obvious correlations of this rapidly varying signal component are visible in the gravitational wave time series.

\begin{figure}[h]
\includegraphics[width=0.55\textwidth]{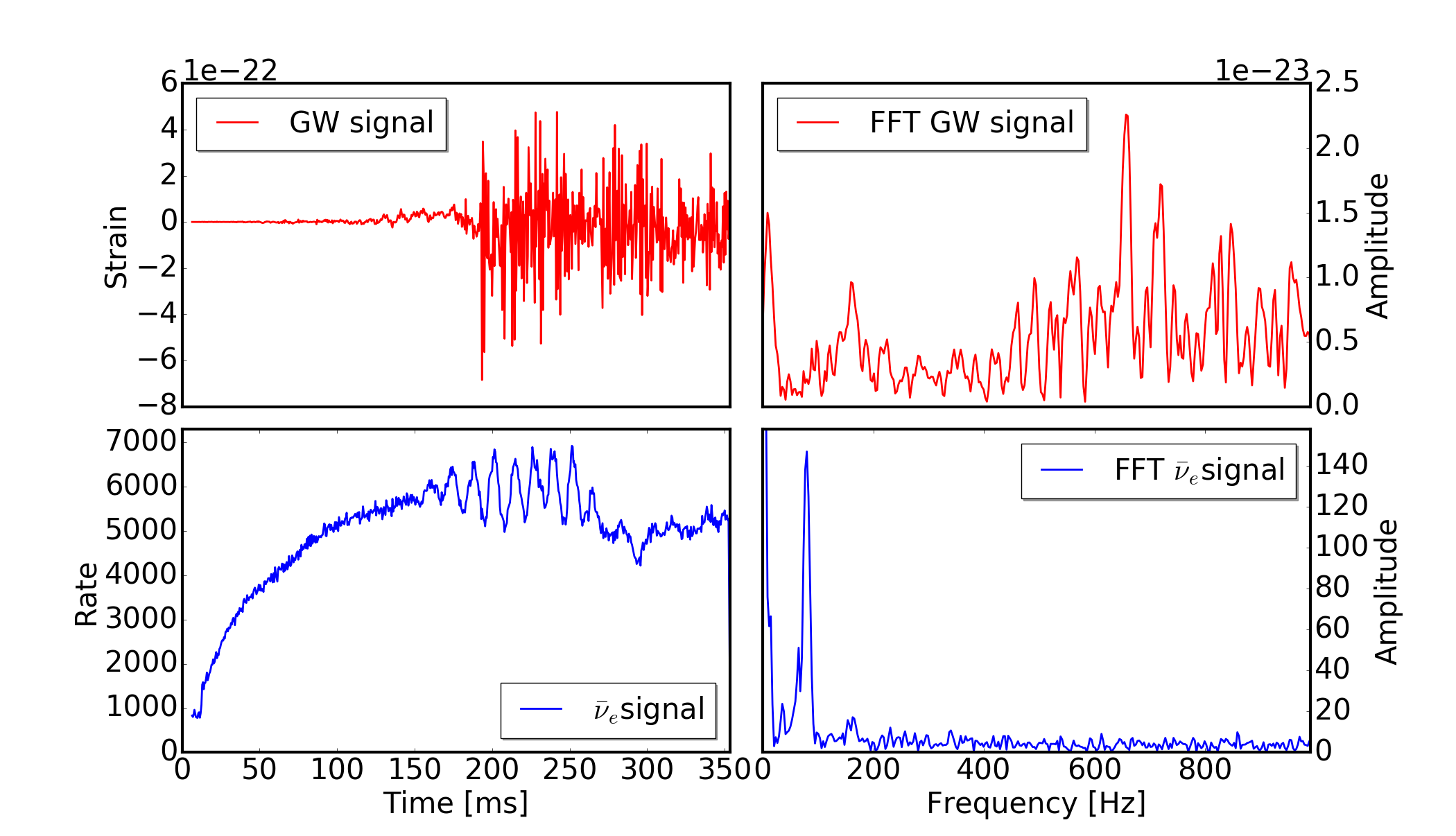}%
\includegraphics[trim=0cm 0cm 0cm 0cm,clip,angle=0,width=0.55\textwidth]{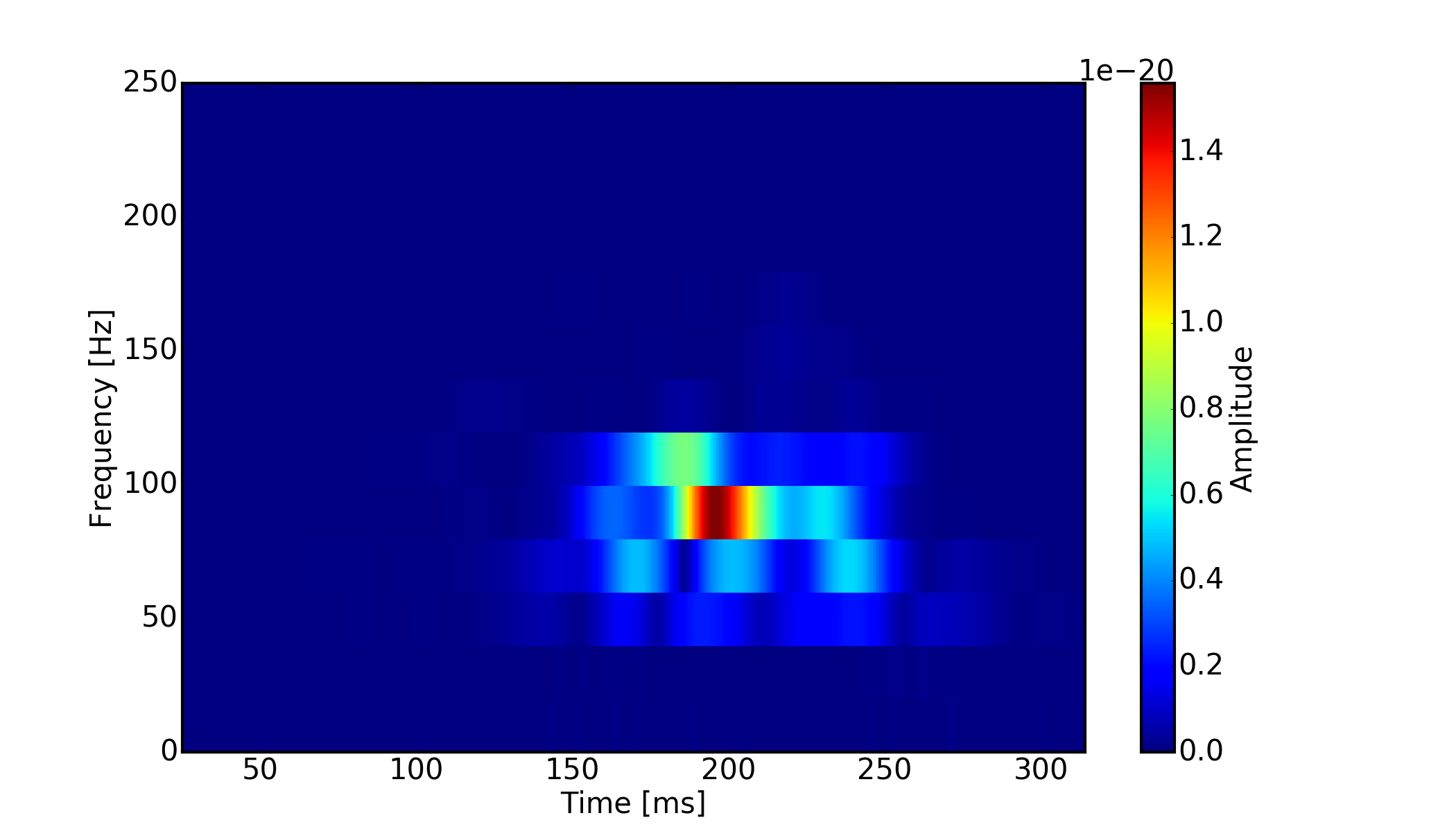}
\caption{\small Left bottom: Expected neutrino signal at 3 kpc distance in IceCube~\cite{bib:Tamborra} (IceCube Monte Carlo), left top: strain amplitude $A+$ of the gravitational wave signal at the source~\cite{bib:Andresen}, warping space time along the vertical and horizontal directions; middle top and bottom: corresponding Fourier spectra. Right: a spectrogram of the cross correlation between neutrino signal frequency and the second harmonic of the gravitational wave signal with aLIGO design detector noise included.}
\label{fig:GW}
\end{figure}

It turns out, however, that the effect of standing wave accretion shocks is present in both signals. The Fourier spectra in Fig.~\ref{fig:GW} (middle) show that the gravitational wave signal, as expected, ocurs with twice the frequency as the neutrino signal. It is also apparent that lower and higher frequencies with larger amplitude contribute to the gravitational wave signal as well.  When performing a sliding window cross correlation analysis between the second harmonic in the neutrino frequency spectrum with the gravitational wave frequency signal in 50 ms time intervals, a clear enhancement emerges around 80~Hz (neutrinos) and 160~Hz (gravitational waves), respectively (Fig.~\ref{fig:GW} (right), with advanced LIGO design detector noise included). This correlation becomes much more apparent at closer distances or when assuming a future gravitational wave detector with lower noise. The higher frequencies seen in gravitational waves do not seem to have obvious counterparts in the simulated neutrino data. 

The possibility to detect oscillations in the 800 Hz range during the first 30 ms for various rotational speeds for rotating progenitors with 12 $M_\odot$ and 40 $M_\odot$~\cite{bib:Ott2} was also investigated. The small signal will only be analyzable for supernovae that are as close as 1 kpc or less~\cite{bib:Natascha}. 
\section{Conclusions and Outlook}
The IceCube observatory provides the world's best statistical accuracy for the neutrino flux of supernovae in our galaxy with a round-the-clock uptime of  currently 99.7\%. However, energies and directions of individual neutrinos can not be determined due to the optical sensor dark rates and -- to a lesser extent -- due to cosmic muons passing the detector. Therefore it is very important to understand the dark rate and cosmic muon characteristics. 
Average neutrino energies can be determined statistically when analyzing the timing information of all hits in the detector. In the future, the sensitivity to the absolute neutrino mass will be assessed, the energy determination will be improved, and the idea of analyzing $ \mathcal{O}(10)\,$s long bursts will be extended to shorter potential signals. To improve the sensitivity of the trigger to a wider range of models as well as unusual hadronic physics or physics beyond the Standard Model, we are implementing a time-domain search using the Bayesian Blocks algorithm~\cite{bib:BayesianBlock}. This technique allows the data themselves to determine the natural timescale of excess counts above background. 
\newpage
\section*{Acknowledgements}
We thank Irene Tamborra, Haakon Andresen and Alexander Summa for their helpfulness and providing access to their simulation data and appreciate suggestions by Sarah Gossan on the implementation of gravitational wave detector noise.

\end{document}